\documentclass[a4paper,11pt]{article}
\usepackage{jinstpub} 
\usepackage{lineno}
\usepackage[version=4]{mhchem}
\usepackage{siunitx}
\usepackage{physics}
\usepackage{orcidlink}

\title{A Simulation Framework for Ramsey Interferometry}

\author[a,1]{Linus B. Persson\note{Corresponding author.}\orcidlink{0009-0008-8835-292X},}
\affiliation[a]{Department of Physics, Lund University,\\
Box 118, 221 00 Lund, Sweden}
\emailAdd{linus.persson@fysik.lu.se}

\author[b]{Peter Fierlinger,}
\affiliation[b]{Physik-Department, Technische Universität München,\\
James-Franck-Str. 1, 857 48 Garching, Germany}

\author[a]{Matthias Holl\orcidlink{0000-0002-7346-4047},}

\author[a,c]{and Valentina Santoro\orcidlink{0000-0001-5379-8771}}
\affiliation[c]{European Spallation Source,\\
Box 176, 221 00, Sweden}

\abstract{The sensitivity of Ramsey interferometry experiments is governed by the interplay between the beam phase-space distribution and the magnetic field environment through which the spins propagate. Quantitative optimisation thus requires a consistent treatment of optics, magnetics and spin dynamics. We present a simulation framework that enables such an analysis by combining neutron optics simulations in \textsc{McStas}, magnetic field modelling in \textsc{COMSOL} and spin-dynamics simulation in the new \textsc{RamseyProp} program. We describe how important experimental parameters such as adiabaticity, flip angle distributions and Ramsey fringe contrast can be studied. The code is being applied to design an experiment to search for axion-like particles at the European Spallation Source (ESS). We examine how the pulsed time structure of the ESS can be exploited to perform Ramsey interferometry on a broad neutron velocity spectrum. In the absence of velocity or timing restrictions, the standard deviation of the spin flip angle at zero detuning can be reduced from 0.67 to 0.17 radians using time-dependent amplitude modulation. Similarly, the phase sensitivity can be improved by a factor $\sim 4$ for a 10~m long setup starting 15~m from the ESS moderator.}


\begin{document}
\maketitle
\flushbottom

\section{Introduction}
Ramsey interferometry is a widely used technique for precision measurements of frequency shifts in quantum systems. Originally developed for molecular beam resonance experiments \cite{Ramsey}, it is now routinely employed in applications ranging from atomic clocks \cite{clock} and magnetometry \cite{magnetometry} to searches for electric dipole moments \cite{EDM} and other signatures of physics beyond the Standard Model. The method relies on separating two coherent interactions by a free precession interval, allowing the system to accumulate a phase that can be converted into a measurable population difference.

In a typical Ramsey beam implementation, a polarised particle beam enters a region with a static magnetic field $B_0$ in the same direction as the incoming polarisation. A first radio-frequency (RF) pulse with amplitude $B_1$ and frequency $\omega_1$ rotates the spins by approximately $\pi/2$, creating a coherent superposition state. The spins then precess freely at the Larmor frequency $\omega_L = \gamma B_0$, where $\gamma$ is the gyromagnetic ratio. After a free precession time $T$, a second RF pulse is applied. In the limit of short pulse durations $\tau \ll T$, the rotations may be treated as instantaneous. Depending on the detuning $\Delta = \omega_1-\omega_L$ and accumulated phase difference $\Delta T$, the spins are either rotated back to their initial orientation ($\Delta T = (2n+1)\pi, n\in \mathbb{Z}$) or inverted $(\Delta T=2n\pi)$, producing the characteristic Ramsey interference pattern when scanning the RF frequency $\omega_1$. Finite pulse durations modify the fringe envelope but do not alter the basic phase sensitivity.

This paper presents a simulation framework for Ramsey beam experiments, allowing the experiments to be modelled using realistic beam phase-space distributions and magnetic field maps. The framework can be used to produce spin flip angle distributions, adiabaticity calculations, Ramsey fringes and sensitivity estimates. This allows for iterative optimisation of optics, magnetostatics and coil systems, supporting design decisions prior to hardware implementation while significantly reducing the reliance on empirical prototyping. 

While the framework is generally applicable, we will here analyse the setup of a proposed search for axion-like particles (ALPs) at the European Spallation Source (ESS) \cite{PRL}. ALPs are well-motivated extensions of the Standard Model that arise in many theories with spontaneously broken global symmetries, making them compelling dark-matter candidates \cite{axionreview, snowmass}. In the ultralight mass range (\hbox{$\ll \SI{}{eV}$}), ALPs behave as a coherently oscillating field that can induce tiny time-dependent energy shifts in spin-polarised systems through the so-called axion wind effect. This manifests as an effective oscillating pseudomagnetic field acting on nucleon spins, making Ramsey interferometry a particularly sensitive probe of the axion-nucleon coupling \cite{theory1, theory2}.

\begin{figure}[t]
  \centering
  \includegraphics[width=1.0\textwidth]{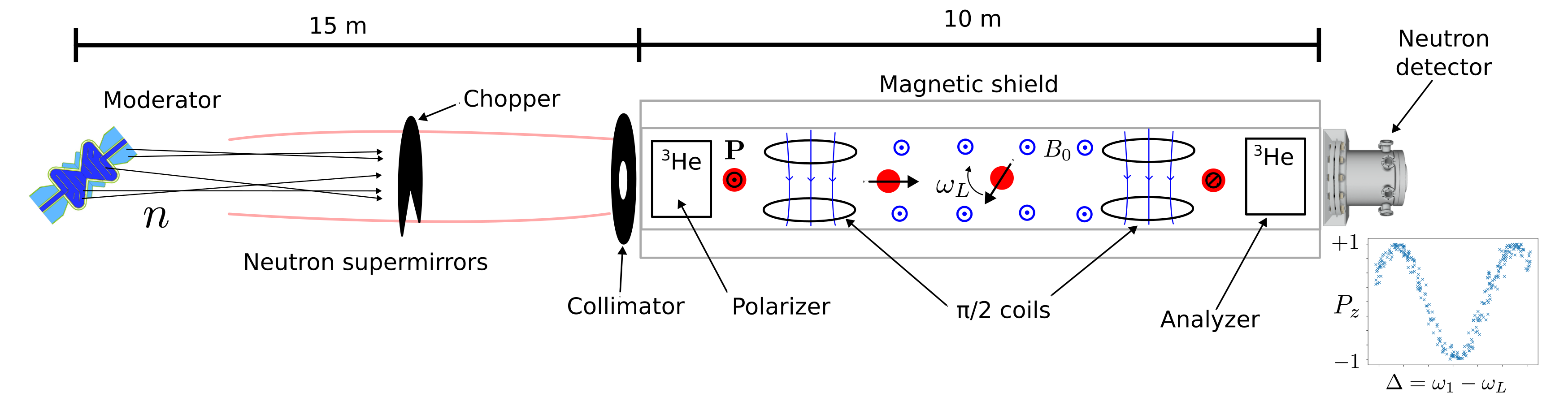}
  \caption{Conceptual sketch of an experiment to search for axion-like particles at the European Spallation Source. Neutrons emanating from the moderator are focussed, chopped and collimated before entering a 10~m long magnetic shield. Inside the magnetic shield is a neutron polarizer, a Ramsey setup with constant $B_0$ field and RF coils, which is followed by an analyser and a neutron detector.}
  \label{Figure1}
\end{figure}

The conceptual setup proposed for the ALP experiment is shown in Fig.~\ref{Figure1}. Neutrons from the ESS moderator are focussed using the high-intensity neutron extraction system designed for the HIBEAM beamline \cite{HIBEAM}. Neutron choppers can be employed to define the temporal structure of the beam and enable time-of-flight determination of the neutron velocity. The beam is collimated before entering a 10~m long magnetically controlled region. The region is protected by a dual-layer octagonal mu-metal shield limiting the external field to $\sim \text{nT}$ level. Inside the shield is a \ce{^3{He}} polarizer, followed by a Ramsey setup. After the second RF coil, a \ce{^3{He}} analyser along with a neutron detector \cite{detector} is used to determine the outgoing polarisation. Since the ESS is a pulsed neutron source, the arrival time of neutrons can be correlated to velocity, even in the absence of neutron choppers. In this paper, we will therefore employ the analysis framework to investigate and quantify the potential of exploiting the pulse structure for velocity-compensated Ramsey interferometry, using time-dependent RF coil amplitude modulation.

\section{Theory}
\subsection{Experimental sensitivity}
In Ramsey interferometry, the measured observable is the outgoing neutron spin polarisation. For an idealised scenario of neutrons with incident polarisation $P=+1$ undergoing instantaneous $\pi/2$ flips, the outgoing polarisation $P$ as a function of detuning $\Delta = \omega_1 - \omega_L$ is given by
\begin{equation}
P(\Delta) = -\cos(\Delta T),
\end{equation}
where $T = L/v$ is the free precession time, $L$ the precession length, and $v$ is the neutron velocity. In a realistic experiment, velocity spread, magnetic-field inhomogeneities, and imperfect pulse areas will reduce the visibility of the Ramsey fringes and introduce offsets. The observed polarisation can then be locally parameterised as
\begin{equation}
P(\Delta) = C \cos(\Delta T + \phi_0) + P_\text{off} ,
\end{equation}
where $C$ is the fringe contrast, $\phi_0$ is a phase, and $P_\text{off}$ accounts for small offsets due to imperfect polarisation and the finite velocity spread of the beam. 

An additional effective pseudomagnetic field $\delta B$, for instance caused by an axion-like particle field (see Refs. \cite{theory1, theory2} for details), contributes to the accumulated phase during the free precession by an amount
\begin{equation}
\delta\phi = \gamma T \, \delta B,
\end{equation}
which manifests as a horizontal shift of the Ramsey fringe. The experimental objective is to determine the phase $\phi_0$, or equivalently the shift in the detuning $\Delta$ with the highest possible precision. Rather than fitting the whole Ramsey fringe, optimal sensitivity to small phase shifts is obtained by operating near the points of steepest slope, defined by where
\begin{equation}
\Delta T + \phi_0 \approx \pm \frac{\pi}{2}.
\end{equation}
This choice maximises the magnitude of $\mathrm{d}P/\mathrm{d}\Delta$, thereby minimising the propagated uncertainty on the inferred detuning shift for a given polarisation uncertainty. Near such a working point $\Delta_0$, the polarisation may be linearised as 
\begin{equation}
P(\Delta) \approx a + b \,(\Delta - \Delta_0),
\end{equation}
with a slope given by
\begin{equation}
b = \left. \dv{P}{\Delta} \right|_{\Delta_0} = \mp C T.
\end{equation}
In practice, the slope $b$ is extracted by measuring the polarisation at several detunings around $\Delta_0$ and performing a weighted linear fit. The uncertainty on the inferred detuning shift from a single linear regression is then given by
\begin{equation} \label{phase}
\sigma_{\Delta} = \frac{\sigma_a}{|b|},
\end{equation}
where $\sigma_a$ is the uncertainty on the fitted polarisation intercept at the working point. For statistically limited measurements, the polarisation uncertainty scales as $\sigma_a \propto 1/\sqrt{N}$, where $N$ is the number of detected neutrons contributing to the signal. Consequently, the detuning uncertainty scales as
\begin{equation}
\sigma_{\Delta} \propto \frac{1}{CT\sqrt{N}} .
\end{equation}
The uncertainty on the inferred effective magnetic field shift is
\begin{equation}
\sigma_{\delta B} = \frac{\sigma_{\Delta}}{\abs{\gamma}}
\propto \frac{1}{\abs{\gamma} C T \sqrt{N}}.
\end{equation}
This expression highlights that the statistical sensitivity of a Ramsey-based axion search is governed by three factors: the fringe contrast $C$, the effective spin precession time $T$, and the total number of detected neutrons $N$. In a pulsed experiment such as at the ESS, the phase can be extracted independently for each neutron pulse, such that the overall sensitivity after a run of given duration scales as the square root of the number of pulses accumulated during that time. This scaling assumes that the axion-induced effective field remains coherent over the integration time.

In the context of ultralight axion-like particles, the effective pseudomagnetic field is expected to oscillate at a frequency determined by the axion mass $m_a$ according to
\begin{equation}
f_a=\frac{m_a c^2}{h}.
\end{equation}
For the measurement strategy considered here, the phase is extracted independently from successive neutron pulses and then tracked as a function of time. The maximum oscillation frequency that can be resolved is therefore set by the ESS pulse repetition frequency of \SI{14}{Hz}. Restricting the analysis to frequencies safely below the Nyquist limit, i.e. to the few-Hz range, corresponds to ALP masses up to approximately $m_a\lesssim \SI{e-14}{eV}/c^2$.

The simulations presented in this work quantify the Ramsey phase sensitivity for a given optics and magnetics configuration. Translating this phase sensitivity into a projected sensitivity to the axion-nucleon coupling $g_{aNN}$ requires several additional assumptions. In particular, the projection depends on the final neutron optics and magnetic-field design, the resulting neutron intensity and velocity distribution, the local dark-matter density, the coherence properties and velocity distribution of the ALP field, the relative orientation of the axion wind and spin quantisation axis, and the statistical procedure used to extract the oscillatory phase signal. A dedicated coupling-sensitivity study lies beyond the scope of the present instrumentation paper, but is planned once the experimental design is completed.

\subsection{Time-dependent amplitude modulation}
The transition probability for a neutron spin to flip inside an RF coil with fixed driving frequency $\omega_1$ and passage time $\tau$ through the coil can be calculated from Rabi’s formula,
\begin{equation}
P_{\uparrow\downarrow}(\Delta) = \frac{\Omega_R^2}{\Omega_R^2 + \Delta^2} \sin^2 \left( \frac{\sqrt{\Omega_R^2+\Delta^2}}{2}\,\tau \right),
\end{equation}
where $\Omega_R=\gamma B_1$ and $\Delta=\omega_1-\omega_L$. In the Bloch-vector picture used in these simulations, this is equivalent to a deterministic spin rotation whose angle $\theta$ is related to the transition probability by $P=\sin^2(\theta/2)$. 

A real neutron beam will contain a spread of incident neutron velocities and corresponding passage times, which significantly deteriorates the fringe contrast, while monochromatising the beam necessarily comes at the cost of reduced intensity. One way of compensating for the velocity spread is by time-dependent amplitude modulation \cite{velocity}. For a coil extending from $z_1$ to $z_2$, and assuming all neutrons are created at $z=0$ when $t=0$, a time-dependent amplitude $f(t)$ can be defined which ensures that the integrated field is the same for all velocities, i.e.
\begin{equation}
\forall v: \theta(v) = |\gamma|B_1 \int_{z_1/v}^{z_2/v} f(t) \mathrm{d}t \equiv \frac{\pi}{2}.
\end{equation}
The function $f(t)$ satisfying this relationship is of the form
\begin{equation} \label{envelope}
f(t)= \frac{\pi}{2 |\gamma| B_1 \ln(z_2/z_1) t}.
\end{equation}
When applying this method to the ESS pulse which is extended with a duration of 2.86 ms, neutrons will on average arrive slightly later (counting $t=0$ as the beginning of the pulse) than the above calculation assumes. By shifting the envelope by approximately half of the pulse width, $t \mapsto t-t_0$ with $t_0=\SI{1.5}{ms}$, this can partly be compensated. It should be noted that the envelope function needs to be reset with a periodicity matching the ESS repetition rate of 14 Hz (with a period of 71 ms). 

\section{Methodology}
\subsection{The \textsc{RamseyProp} program}
To study the experimental impact of various input particle trajectories and magnetic field distributions, a new Python-based simulation code called \textsc{RamseyProp} has been developed. General neutron ray-tracing packages such as McStas \cite{McStas} and VITESS \cite{vitess} already include components for polarised-neutron transport and spin precession in magnetic fields. \textsc{RamseyProp} extends these capabilities toward dedicated simulations of Ramsey beam experiments by combining event-by-event phase-space information from neutron transport simulations with realistic magnetic field distributions and directly evaluating Ramsey-specific observables such as flip-angle distributions, adiabaticity, Ramsey fringe contrast and phase sensitivity.

The primary input of the code is an MCPL \cite{MCPL} file, a standardised format for storing event-by-event particle data generated by Monte Carlo transport simulations such as \textsc{McStas} \cite{McStas}, VITESS \cite{vitess} or \textsc{PHITS} \cite{PHITS}. The file contains the initial position, energy, velocity, time, and statistical weight of each particle. Acceptance cuts can be applied based on velocity, divergence, and whether the trajectory intersects the defined detector region. Optional time-gating choppers can be implemented, in which case their transmission function at the neutron’s arrival time is evaluated at a specified longitudinal plane and the resulting transmission probability is multiplied by the particle weight. From the surviving particles, a fixed number is selected using weighted reservoir sampling, with selection probability proportional to the effective statistical weight (including the chopper transmission). The propagated neutrons therefore represent an unbiased finite representation of the transmitted beam, while avoiding the need to simulate the full MCPL population. The ability to parametrically scan the transmission window duration for chopper optimisation is also implemented.

\begin{figure}[htbp]
  \centering
  \includegraphics[width=0.75\textwidth]{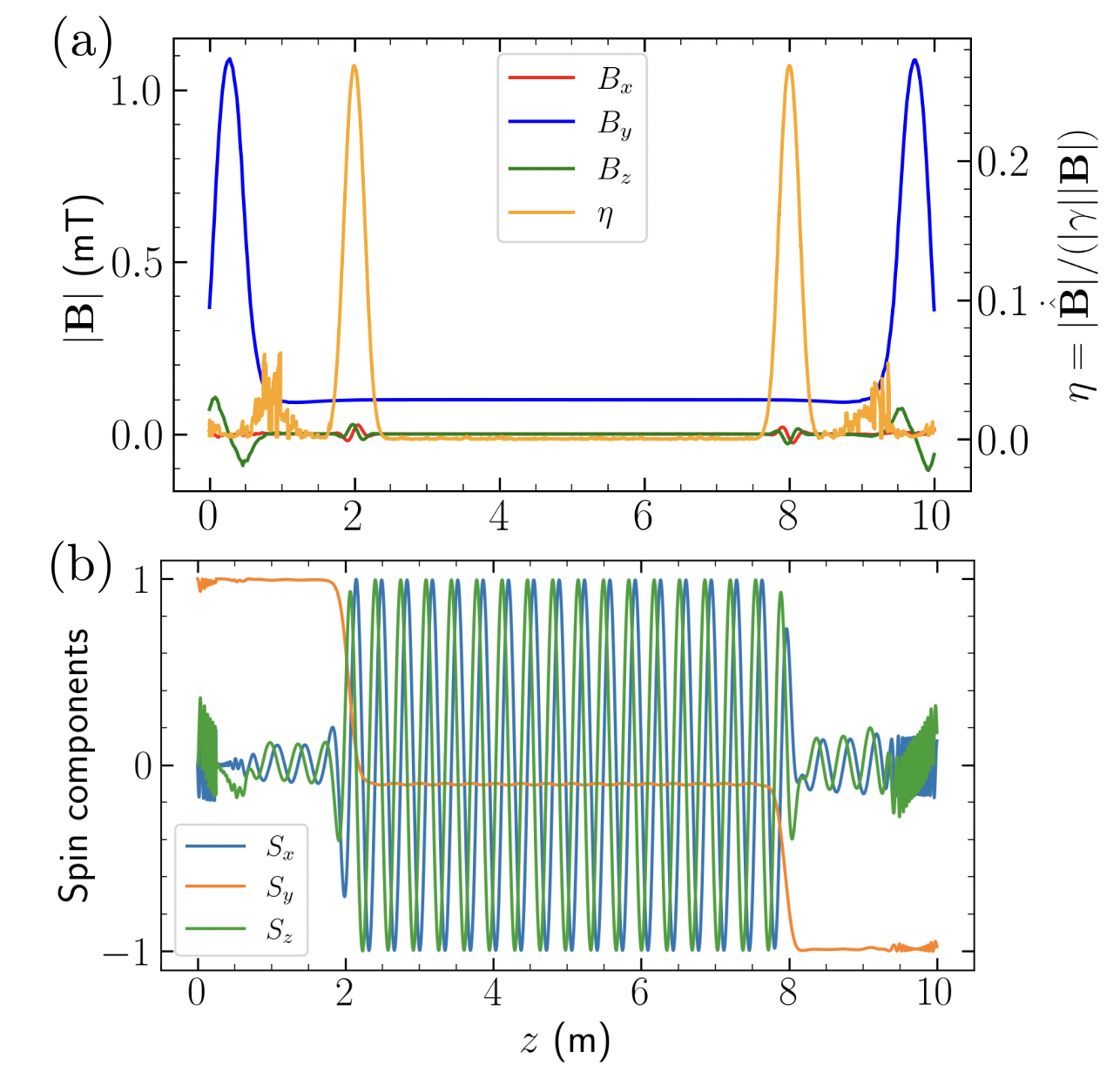}
  \caption{Example outputs from \textsc{RamseyProp}. In this simulation, the magnetic field distributions are interpolated from COMSOL, with an applied $B_0=\SI{100}{\micro\tesla}$ in the central region, higher fields exceeding $\SI{1}{mT}$ close to the edges and an effective $B_1=\SI{28.6}{\micro\tesla}$ from two coils located at $z=\SI{2}{m}$ and $z=\SI{8}{m}$. This particular trajectory has zero detuning, hence an outgoing polarisation of $-1$ is achieved after the second coil. Panel~(a) shows the magnetic field distribution in Cartesian components and the adiabaticity experienced along the trajectory. Panel~(b) shows the corresponding evolution of the spin components as calculated from the Bloch equation. }
  \label{Figure2}
\end{figure}

\textsc{RamseyProp} combines the beam phase space information with magnetic field distributions for the $B_0$ and $B_1$-fields, which can either be defined analytically or be interpolated from the output of \textsc{COMSOL} \cite{COMSOL} simulations into a time-dependent total field $\mathbf{B}(\mathbf{r},t)$. The program can either run with a fixed detuning or change the detuning (and thus the coil frequency) between each simulated particle. The neutrons are transported along their trajectory $\mathbf{r}(t)$, while their spin evolution $\mathbf{S}(t)$ is determined by the Bloch equation
\begin{equation}
\dv{\mathbf{S}(t)}{t} = \gamma \mathbf{S}(t) \times \mathbf{B}(\mathbf{r}(t),t).
\end{equation}
The calculation is performed numerically using a 4th-order Runge-Kutta method and accelerated using the Numba just-in-time compiler \cite{Numba}. In each time step, the magnetic field is interpolated trilinearly. 

Several features are implemented to facilitate Ramsey interferometry simulations in particular. Polariser (where the polarisation direction is set) and analyser (where the projection of the polarisation on the quantisation axis is measured) positions can be defined anywhere along the trajectory. Furthermore, the detector plane can be partitioned into arbitrary user-defined spatial regions, and the spin evolution can be analysed separately for each region via independent simulation runs. Time-dependent amplitude modulation may also be defined for each coil in the simulation. The code has the ability to plot the magnetic field experienced by each neutron along its trajectory and the corresponding adiabaticity
\begin{equation}
\eta(t)=\frac{|\mathrm{d}\hat{\mathbf{B}}/\mathrm{d}t|}{|\gamma|\,|\mathbf{B}|}, \quad \hat{\mathbf{B}} = \frac{\mathbf{B}}{|\mathbf{B}|}.
\end{equation}
Examples of such plots are shown in figure~\ref{Figure2}.   

\begin{figure}[htbp]
  \centering
  \includegraphics[width=0.65\textwidth]{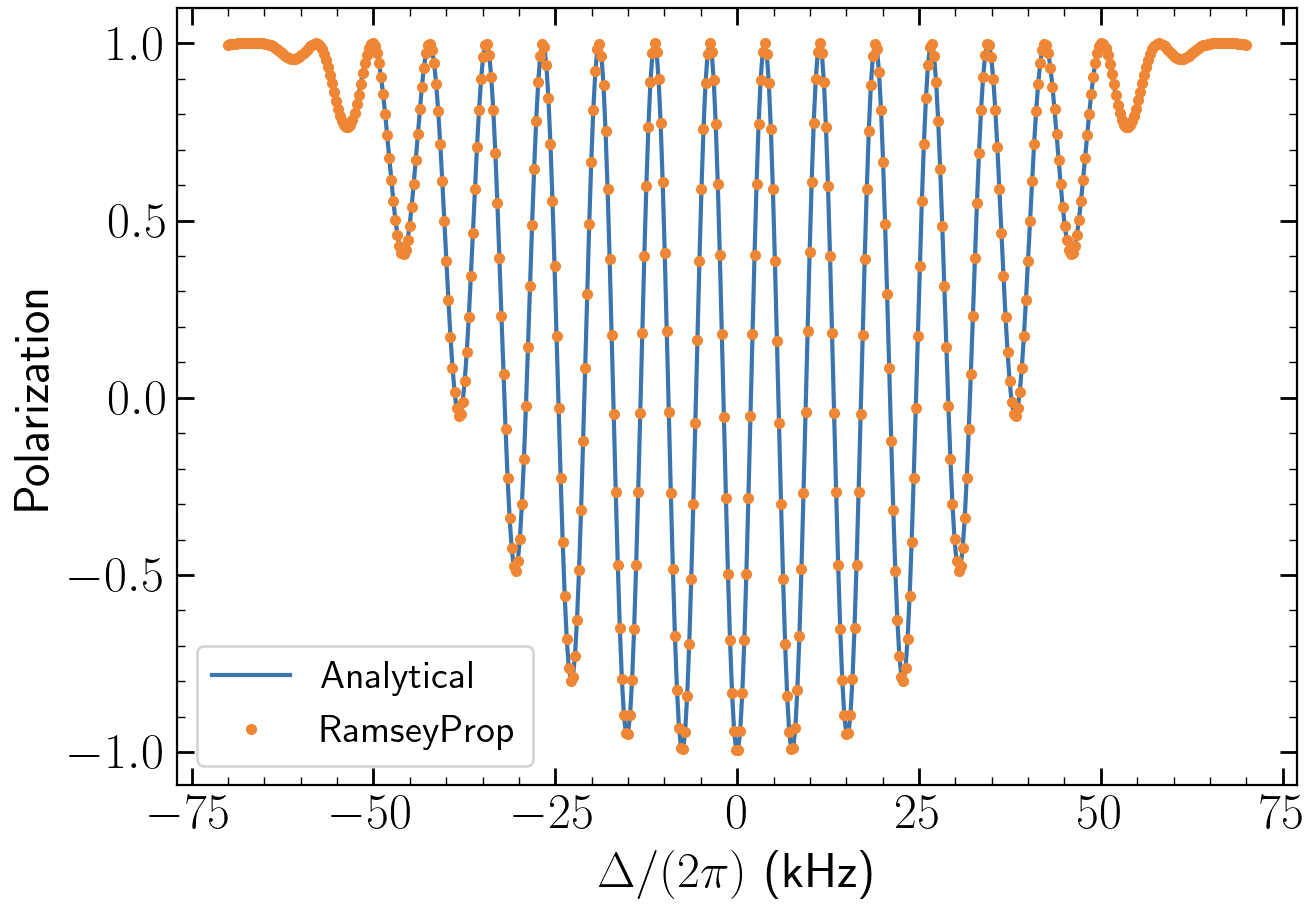}
  \caption{The Ramsey fringes obtained for an interferometry setup with free precession length $L=\SI{8}{cm}$, coil length $\ell=\SI{1}{cm}$, $B_0=\SI{10}{mT}$, and $B_1=\SI{600}{\micro\tesla}$ for a single neutron velocity $v=\SI{700}{m/s}$. The blue line indicates the analytical Ramsey equation while the 500 orange points correspond to individual neutrons simulated in \textsc{RamseyProp}.}
  \label{Figure3}
\end{figure}

To assess the performance of the code, we consider a simple configuration with fixed neutron velocity $v=\SI{700}{m/s}$, free precession length $L=\SI{8}{cm}$, coil length $\ell=\SI{1}{cm}$, $B_0=\SI{10}{mT}$ and $B_1=\SI{600}{\micro\tesla}$. Under these conditions, the Ramsey equation can be used to calculate the fringes analytically. A root mean square error of $\SI{5e-4}{}$ is achieved when simulated in \textsc{RamseyProp} using a time step of $\text{d}t=\SI{5e-8}{s}$, confirming the validity of the simulation. Such a simulation with 500 trajectories takes on the order of a few minutes on standard commercial hardware. A comparison of the analytical solution and the simulation output is shown in figure~\ref{Figure3}.

\subsection{Simulation inputs}
\begin{figure}[htbp]
  \centering
  \includegraphics[width=1.0\textwidth]{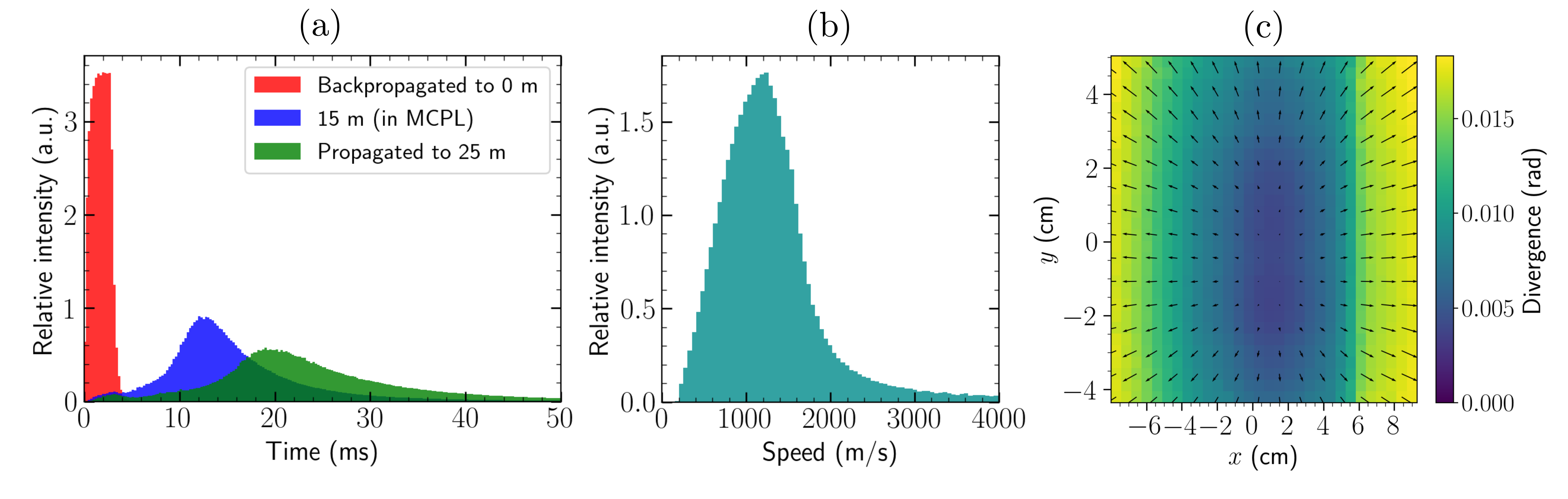}
  \caption{Spectral distributions of the HIBEAM beamline at the ESS E5 beam port as used for simulation in \textsc{RamseyProp}. Panel~(a) shows the time distribution in the MCPL file (15 m from the moderator) as well as the extrapolated distributions at the moderator and at 25~m. Panel~(b) shows the velocity distribution and Panel~(c) shows the divergence distribution relative to the longitudinal axis of the Ramsey setup.}
  \label{Figure4}
\end{figure}

The simulations presented here are based on the neutron spectrum of the E5 beam port at the European Spallation Source, the location of the proposed HIBEAM particle physics beamline. The beamline is simulated in the neutron ray-tracing simulation package \textsc{McStas} \cite{McStas} using a model of the neutron extraction system installed in the ESS monolith. The simulation assumes an elliptical $m=3$ guide outside the monolith, consistent with the geometry presented in Ref.~\cite{HIBEAM}. A set of slits located around 11.5~m from the moderator collimates the beam. The particles are sampled at the exit of the ESS bunker, a concrete shielding structure ending 15~m from the moderator. The time, velocity and divergence distributions of the beam at this point are displayed in figure~\ref{Figure4} and used as input for \textsc{RamseyProp}. The time distribution histogram also shows the time distribution at the moderator and at 25~m as calculated by adding or subtracting the distance divided by the current longitudinal velocity. It is observed that the width of the time distribution at the moderator matches the expected ESS pulse width (about 3~ms).

To decouple this methodological study from geometrical specifics relating to the magnetic shielding and coil system, the simulations presented here use analytically defined magnetic fields. A constant magnetic field $B_0=\SI{100}{\micro\tesla}$ is applied in the free precession region and inside the coils. An RF coil amplitude of $B_1 = 34.4 \,\mu\mathrm{T}$ is used, corresponding to $B_1/B_0 \approx 0.3$. In the simulations presented here, the transverse RF field is implemented as a circularly polarised field, such that the rotating-wave approximation is not invoked. This avoids sensitivity to the phase at which the neutron enters the RF region.

\section{Results}

\begin{figure}[htbp]
  \centering
  \includegraphics[width=1.0\textwidth]{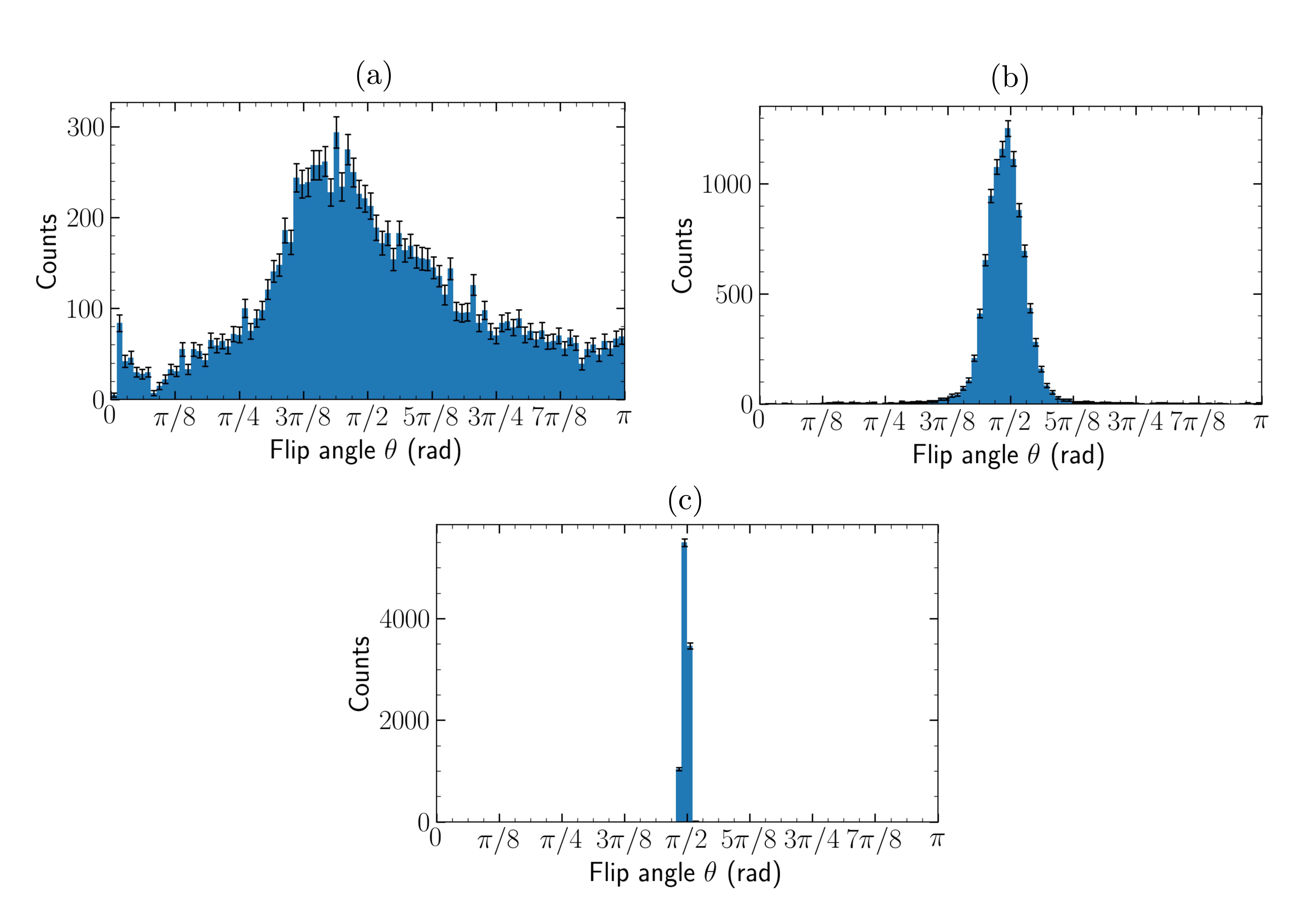}
  \caption{The distribution of flip angles, defined as the arccosine of the projection on the direction of incident spin, after passage through a 0.3~m coil located 16~m from the moderator. Panel~(a) shows the distribution for the full spectrum without velocity compensation or acceptance cuts, while Panel~(b) is the same result when including a time-dependent amplitude modulation of the $B_1$ amplitude. Panel~(c) places additional time restrictions on the beam, assuming passage through a 0.3~ms window at 6.7~m from the moderator. In each case, 10~000 neutrons are simulated at $\mathrm{d}t = \SI{e-6}{s}$.}
  \label{Figure5}
\end{figure}

Consider the case when the full neutron spectrum of the HIBEAM beamline, without any time or velocity restrictions, is sent into an RF coil located \SI{16}{m} from the moderator. The coil has length \SI{0.3}{m} and field strength $B_1=\SI{34.4}{\micro\tesla}$. The spin flip angle distribution at zero detuning is shown in figure~\hyperref[Figure5]{\ref*{Figure5}a}. While the distribution has an average of about $\pi/2$ (the $B_1$ value was chosen here to match a neutron velocity of $v=\SI{1200}{m/s}$), the spread is very large with a standard deviation of 0.67 radians. In figure~\hyperref[Figure5]{\ref*{Figure5}b}, time-dependent amplitude modulation is applied using the envelope function in eq. \eqref{envelope}. The distribution becomes narrower around $\pi/2$ with a standard deviation of 0.17 radians. In figure~\hyperref[Figure5]{\ref*{Figure5}c}, we only consider neutrons which pass through a 0.3~ms time window (centred at 7.08~ms) 6.7~m from the moderator, which is the proposed location of a pulse definition chopper. We may consider this location as the starting point of the envelope function, i.e. by setting $t_0=\SI{7.08}{ms}$. The distribution then becomes as shown in figure~\hyperref[Figure5]{\ref*{Figure5}c}. This reduces the standard deviation further to 0.02 radians, while eliminating the wide wings of the distribution present in figure~\hyperref[Figure5]{\ref*{Figure5}b}. 

\begin{figure}[htbp]
  \centering
  \includegraphics[width=0.95\textwidth]{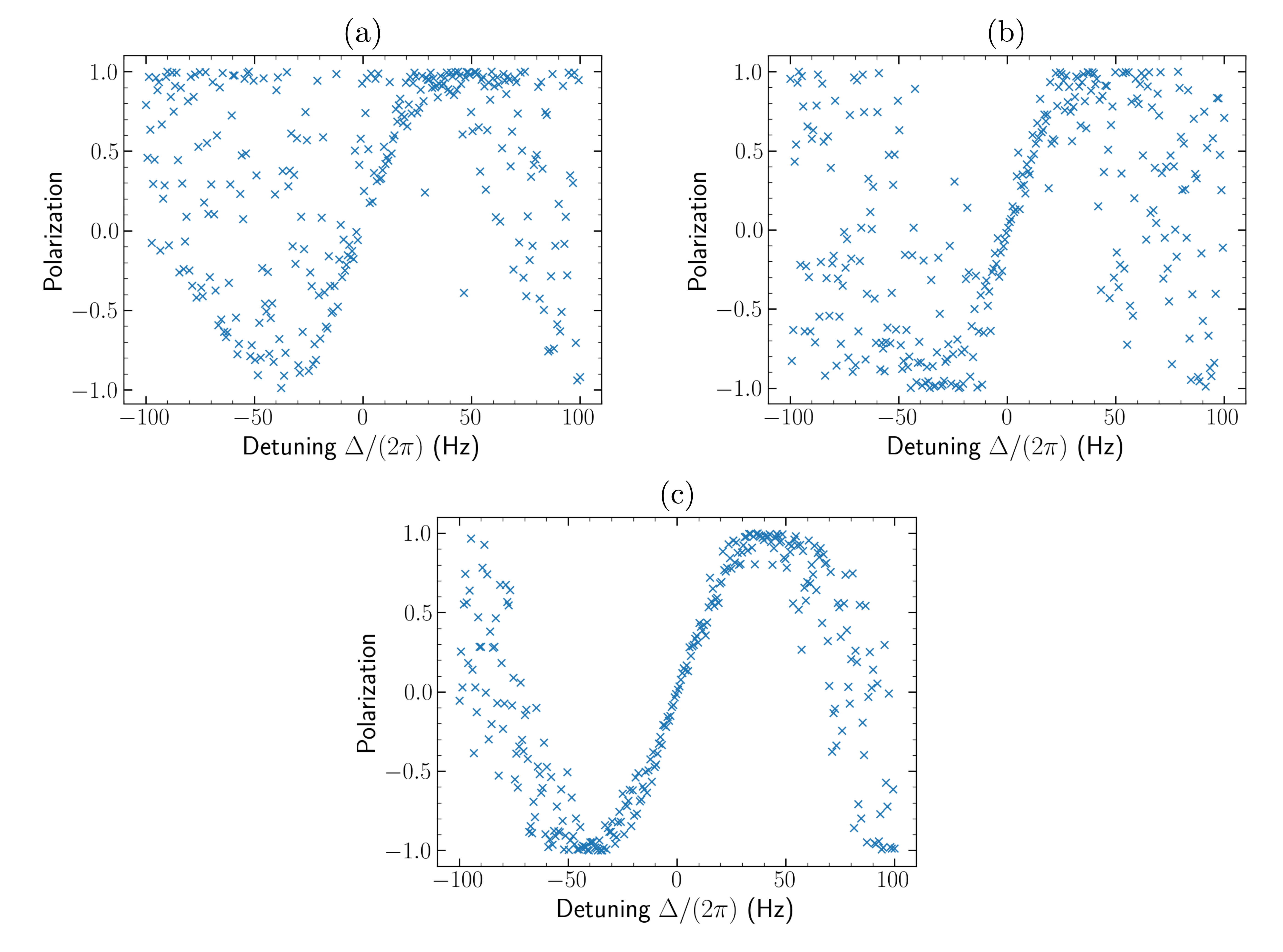}
  \caption{Ramsey fringes obtained for neutrons from the proposed HIBEAM beamline at the E5 beam port. Each mark corresponds to a simulated neutron trajectory at $\mathrm{d}t = \SI{e-6}{s}$. Panel~(a) shows the fringes for the full velocity spectrum without velocity compensation or acceptance cuts, while Panel~(b) is the same result when including a time-dependent amplitude modulation of the $B_1$ amplitude. Panel~(c) places additional time restrictions on the beam, assuming passage through a 0.3~ms window at 6.7~m from the moderator.}
  \label{Figure6}
\end{figure}

We now extend this simulation by including a 7.7~m long free precession length between two coils. Here we shift the Ramsey fringe to a working point on the steepest slope by changing the relative phase of the second coil by $\pi/2$. The resulting Ramsey fringes for each of the three aforementioned cases are shown in figure~\ref{Figure6}. Each cross corresponds to a simulated neutron trajectory, with a total of 300 neutrons simulated in each case. A clear improvement in contrast is observed near the working point when introducing the time-dependent amplitude modulation. While the effect deteriorates at larger detunings, the phase information is extracted close to the working point, and so does not invalidate the method. When timing restrictions are applied, the fringe contrast improves across a significantly larger detuning range.

\begin{figure}[htbp]
  \centering
  \includegraphics[width=0.68\columnwidth]{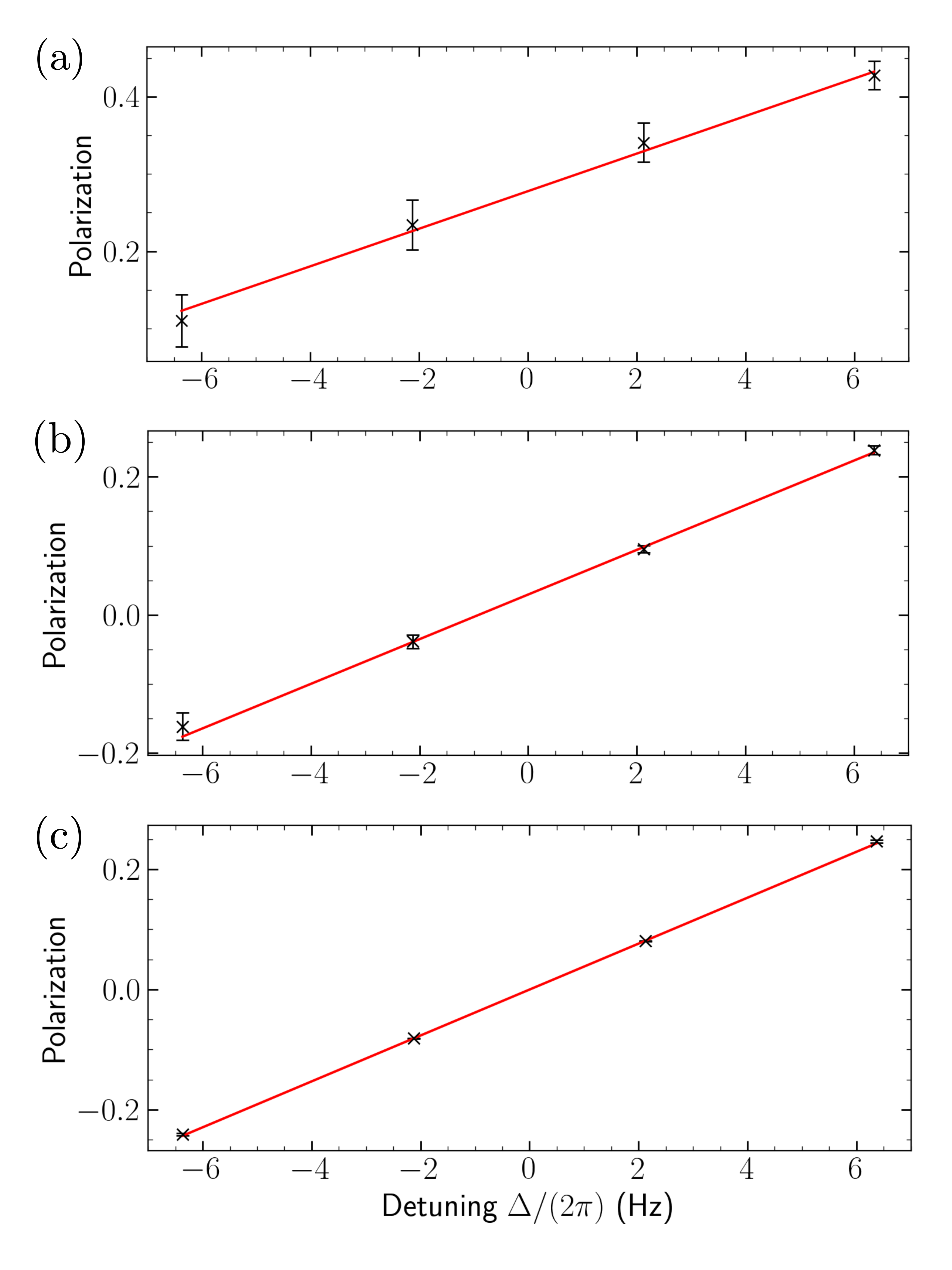}
  \caption{Linear fits to the steepest slope of the Ramsey fringes. Each point corresponds to the simulation of 250 neutron trajectories at $\mathrm{d}t = \SI{e-6}{s}$. Panel~(a) shows the fit for the full velocity spectrum without velocity compensation or acceptance cuts, while Panel~(b) is the same result when including a time-dependent amplitude modulation of the $B_1$ amplitude. Panel~(c) places additional time restrictions on the beam, assuming passage through a 0.3 ms window at 6.7~m from the moderator.}
  \label{Figure7}
\end{figure}

To quantify these observations, figure~\ref{Figure7} shows linear fits around the working point. At each of the four chosen detunings, 250 trajectories are simulated and the mean and standard deviation of the polarisations are used for the linear regression. The phase uncertainty, as computed from eq.~\eqref{phase}, is found to be $\SI{0.6}{Hz}$ without velocity compensation for this simulation. With the amplitude modulation, the phase uncertainty decreases to $\SI{0.14}{Hz}$, equivalent to a factor $\sim 4$ gain in sensitivity. With the timing restrictions, the phase uncertainty is further decreased to \SI{0.013}{Hz}, representing a factor $\sim 44$ relative to the baseline.

\section{Conclusions}
We have demonstrated that the new simulation code \textsc{RamseyProp} enables quantitative optimisation of neutron Ramsey interferometry experiments. In particular, we have shown how spin flip angle distributions, adiabaticity and Ramsey fringe contrast can be studied under varying conditions on the incoming neutron distribution and coil amplitudes. 

We have also illustrated through a realistic simulation how a Ramsey interferometry experiment with a broad velocity spectrum can make use of the ESS pulse structure. Although the 2.86 ms ESS pulse length will limit the achievable performance, a simple $\propto 1/t$ envelope can still improve the sensitivity by about a factor ~4 without any neutron choppers, as demonstrated through steepest-slope linear regression of the Ramsey fringes. Significant increases are possible by chopping the beam to restrict the time and velocity distribution. While this can improve the phase uncertainty, it comes at the cost of reduced intensity, making detailed experimental optimisation necessary. 

The simulation framework is currently being used for magnetics and optics optimisation of the ALP experiment at the ESS. Future work will include a detailed statistical analysis to translate the phase uncertainty into constraints on the axion-nucleon coupling constant. Although the simulations presented here consider the standard Ramsey configuration, the framework also allows the inclusion of intermediate $\pi$ pulses, enabling partial $T_2$ recovery by correcting for spatially symmetric field inhomogeneities and periodic distortions in time.

The source code is available in Ref. \cite{RamseyProp} and licensed under the GNU General Public License v3.0. The McStas models and MCPL files used for the simulations presented in this article are available upon request.

\acknowledgments

The authors gratefully acknowledge support from the Swedish Foundation for Strategic Research under the grant RIF21-0057 (Development of a magnetic control beamline at the ESS) and from the Swedish Research Council under the grant 2024-05656 (The first particle physics experiment at ESS: Search for axion-like particles at the HIBEAM beamline). We further thank Douglas H. Beck at the University of Illinois for useful discussions. During the preparation of this work the authors used GitHub Copilot in order to debug and validate the implemented algorithms. After using these services, the authors reviewed and edited the content as needed and take full responsibility for the content of the published article.


\bibliographystyle{JHEP}
\bibliography{ref.bib}


\end{document}